\documentclass{PoS}
\usepackage[utf8]{inputenc}
\usepackage{amsmath}
\usepackage{amsfonts}
\usepackage{amssymb}
\usepackage{graphicx}
\title{Centre Vortex Effects on the Overlap Quark Propagator}
\ShortTitle{Centre Vortex Effects on the Overlap Quark Propagator}
\FullConference{The 32nd International Symposium on Lattice Field Theory \\
23-28 June, 2014 \\
Columbia University New York, NY}
\author{\speaker{D.~Trewartha}, W.~Kamleh and D.~Leinweber \\
Special Research Centre for the Subatomic Structure of Matter (CSSM) \\
University of Adelaide, Australia}
\abstract{We investigate the role of centre vortices in dynamical mass generation using overlap fermions. The exact chiral symmetry that the overlap fermion action possesses yields a distinctive response to the underlying topology of the gauge field, leading to novel results. We study the quark propagator and associated mass function on gauge field backgrounds featuring the removal of centre vortices as well as on vortex-only backgrounds. The effect of cooling vortex-only backgrounds on the overlap quark propagator is also presented.}
\begin{document}
\newlength{\PlotHeight}
\setlength{\PlotHeight}{60mm}
\section{Introduction}
QCD at non-perturbative energies has two key features; dynamical chiral symmetry breaking and confinement. The underlying mechanism responsible for these, as well as whether they are attributable to a single phenomenon, has been a source of debate. In $\mathrm{SU(2)}$, both confinement \cite{Del Debbio:1996mh,Del Debbio:1998uu,Greensite:2003bk} and dynamical chiral symmetry breaking \cite{de Forcrand:1999ms,Langfeld:2003ev,Gattnar:2004gx} can be attributed to the topological defects associated with the centre degree of freedom of $\mathrm{SU(2)}$, centre vortices. In $\mathrm{SU}(3)$, centre vortices remain closely associated with confinement \cite{Langfeld:2003ev,Bowman:2008qd}, however studies of dynamical chiral symmetry breaking using the AsqTad \cite{Orginos:1999cr} quark propagator have been unable to show a corresponding result \cite{Bowman:2010zr}. \par
We study dynamical chiral symmetry breaking using the superior chiral properties of the overlap quark propagator \cite{Narayanan:1992wx}, and find that we are able to show a loss of dynamical mass generation and thus dynamical chiral symmetry breaking with vortex removal, and reproduce dynamical mass generation on vortex only configurations after cooling.
\section{Centre Vortices}
We will identify centre vortices on the lattice in the standard way using gauge fixing and centre projection. The centre of $\mathrm{SU}(3)$ is given by elements
\begin{equation}
Z_{\mu}(x) = \exp{\bigg [}\frac{2\pi i}{3} m_{\mu}(x){\bigg ]}\mathrm{I}\mathrm{,}\quad m_{\mu} \in \{-1,0,1\}.
\end{equation}
Centre vortices are then the defects in the centre-projected gauge field. \par
\begin{figure}
\begin{center}
\includegraphics[height=80mm,angle=90]{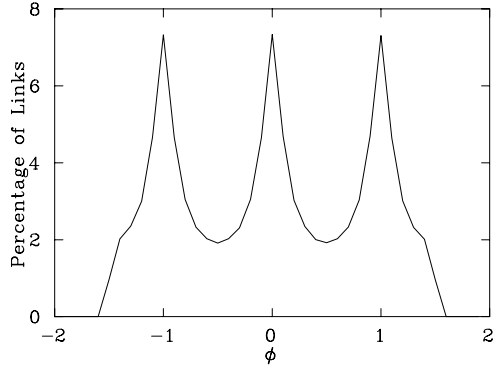}
\caption{Histogram of phases, $\phi$, of traces of links after gauge-fixing to Maximal Centre Gauge}
\label{fig:phihist}
\end{center}
\end{figure}

There are multiple choices of gauge for identifying centre vortices. We use the Maximal Centre Gauge	\cite{Del Debbio:1996mh},
\begin{equation}
R_{mes} = \sum_{x,\mu} |\mathrm{Tr}\,U^{\Omega}_{\mu}(x)|^{2} \rightarrow \, \mathrm{Max},
\end{equation}
to bring links close to centre elements. Then we project onto $Z_{3}$; writing the trace of each link as
\begin{equation}
\frac{1}{3}\mathrm{Tr}\,U_{\mu}^{\Omega}(x) = r_{\mu}(x)\exp(i\phi_{\mu}(x)),
\end{equation}
where we choose $m_{\mu}(x) \in \{-1,0,1\}$ with $\frac{2\pi m_{\mu}(x)}{3}$ closest to $\phi_{\mu}(x)$. A histogram of values of $\phi_{\mu}(x)$ on a typical configuration is shown in Fig.~\ref{fig:phihist} before and after gauge-fixing to Maximal Centre Gauge, showing 3 distinct peaks corresponding to the three centre elements.
The vortex-only theory is then defined using configurations
\begin{equation}
Z_{\mu}(x) = \exp{\bigg [}\frac{2\pi i}{3} m_{\mu}(x){\bigg ]}\mathrm{I},
\end{equation}
and the vortex removed by
\begin{equation}
R_{\mu}(x)=Z^{\dagger}_{\mu}(x)U_{\mu}^{\Omega}(x).
\end{equation}
These configurations form the basis for our analysis, alongside the original configurations, which we denote as 'untouched'.
\section{The Overlap Quark Propagator}
The massive overlap-Dirac operator \cite{Narayanan:1992wx} is given by
\begin{equation}
\label{massov}
D_{o}(\mu) = \frac{(1-\mu)}{2}[1+\gamma_{5}\epsilon{\big (}\gamma_{5}D(m_{w}){\big )}] + \mu ,
\end{equation}
where we use the FLIC operator \cite{Zanotti:2001yb} as overlap kernel $D(m_{w})$, and the value of the overlap mass parameter, $\mu\, =\, 0.022$, corresponding to a physical bare quark mass of $70$ MeV. \newline
The overlap operator has a lattice-remnant chiral symmetry, which allows us to study chirally sensitive properties such as dynamical mass generation. \par
In a covariant gauge, the quark propagator on the lattice has as its general form
\begin{equation}
\label{propgenform}
S(p) = \frac{Z(p)}{iq\!\!\!\!/\ + M(p)},
\end{equation}
where $M(p)$ is the non-perturbative mass function and $Z(p)$ contains all renormalisation information, allowing us to study the presence of dynamical chiral symmetry breaking through the infrared behaviour of the mass function. In our analysis, we use a cylinder cut on this data \cite{Leinweber:1998im}, and renormalise $Z(p)$ to be $1$ at the largest momentum value considered.
\section{The Overlap Quark Propagator on Vortex Only and Vortex Removed backgrounds}
Results are calculated on $50$ $20^{3} \times 40$ gauge-field configurations using Lu\"scher-Weisz $\mathcal{O}(a^{2})$ mean-field improved action \cite{Luscher:1984xn} with a lattice spacing of $0.125 \, \mathrm{fm}$. We fix to Landau gauge using a Fourier transform accelerated algorithm \cite{Davies:1987vs} to the $\mathcal{O}(a^2)$ improved gauge-fixing functional \cite{Bonnet:1999mj}. \par
\begin{figure}
\includegraphics[height=\PlotHeight]{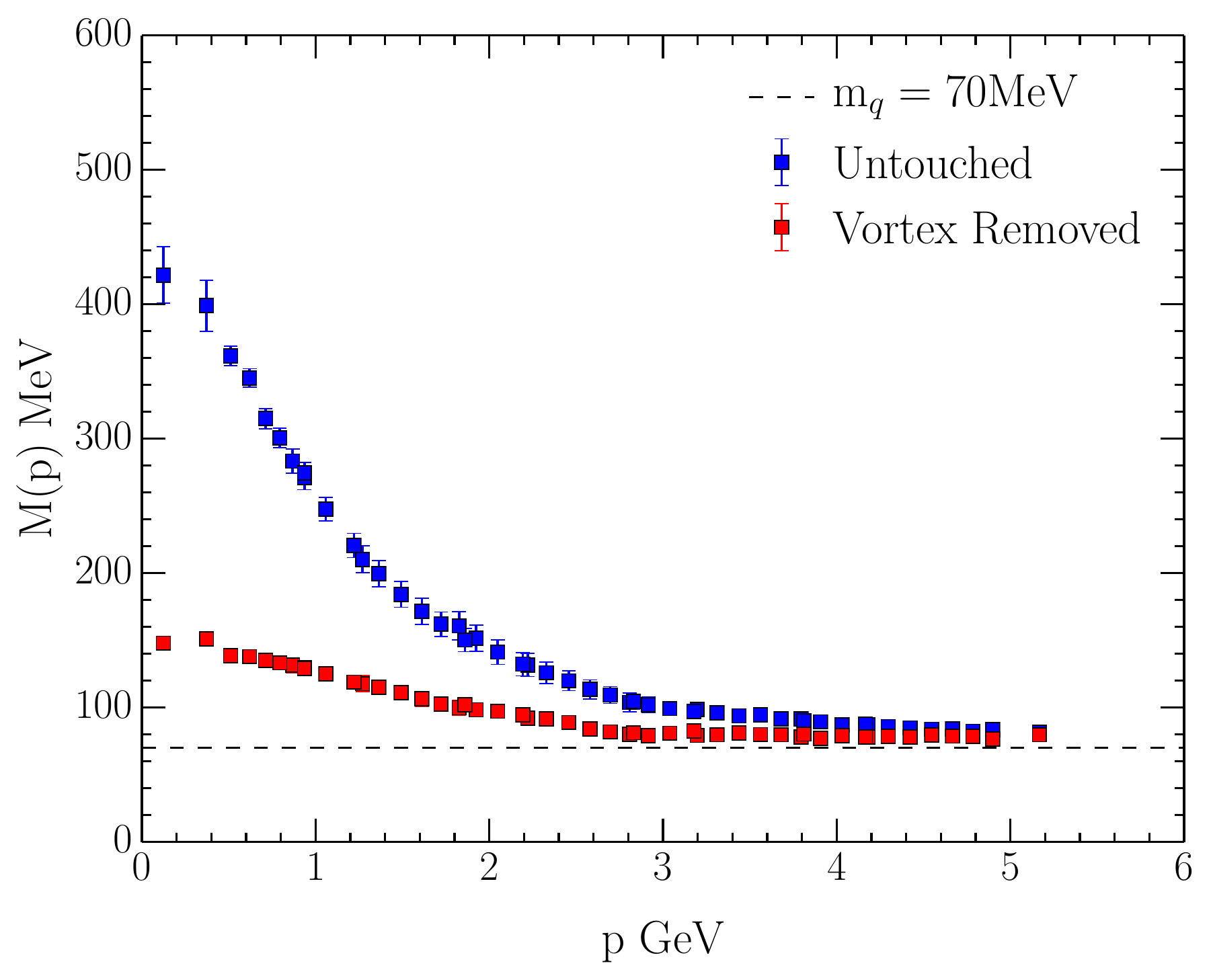}
\includegraphics[height=\PlotHeight]{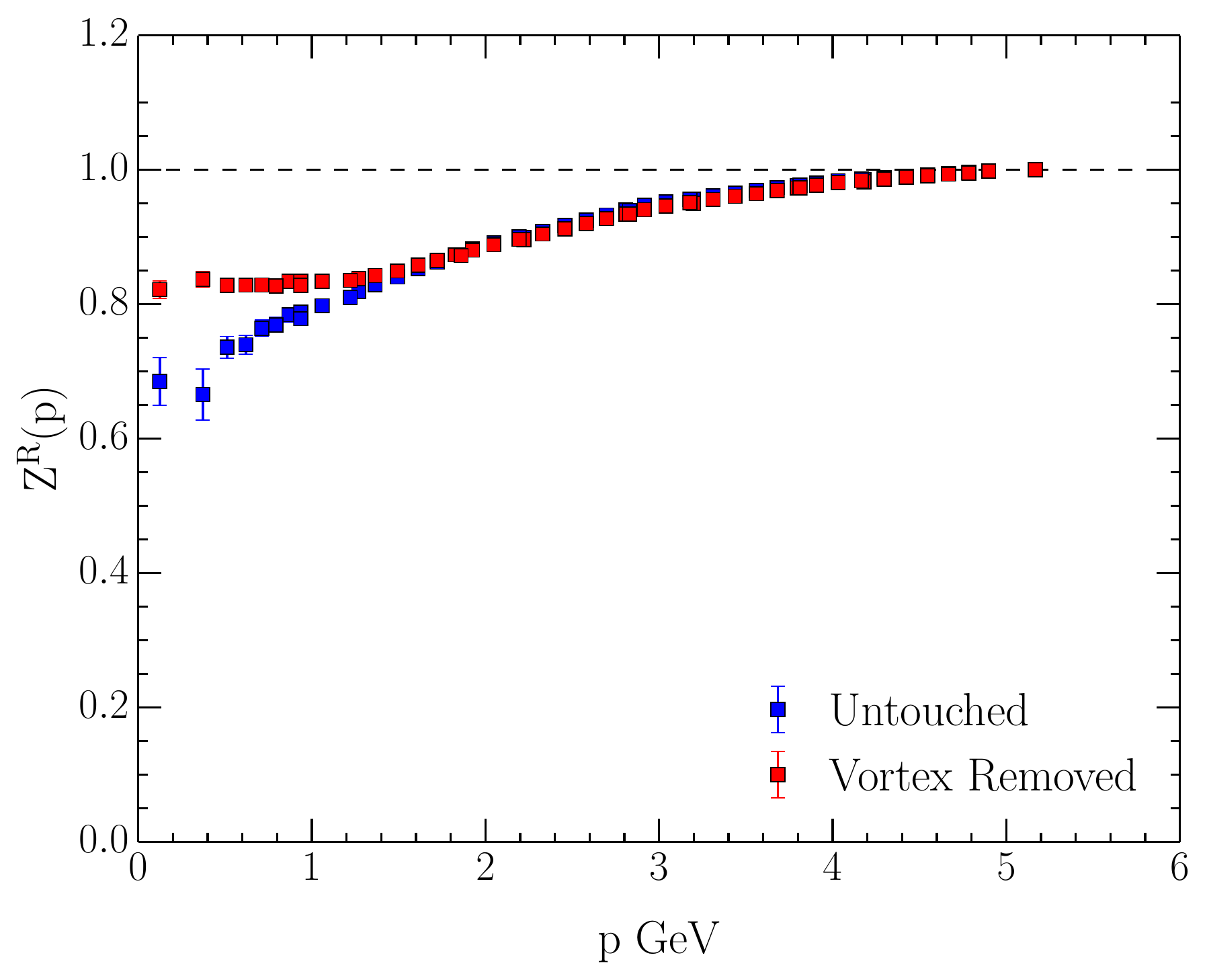}
\caption{The mass and renormalisation functions on original (untouched) and vortex removed configurations, at an input bare quark mass of $70$ MeV}
\label{fig:M02200UTVR}
\end{figure}
The mass and renormalisation functions for untouched and vortex removed configurations are shown in figure \ref{fig:M02200UTVR}. The untouched results reproduce the well-known shapes of these functions; the mass function is strongly enhanced in the infrared by dynamical chiral symmetry breaking, tailing off in the ultraviolet. The renormalisation function is mildly suppressed in the infrared, and reaches a plateau in the ultraviolet limit. \par
By contrast, the vortex removed configurations are far more tree-like. The mass function has very little enhancement in the infrared, showing the removal of dynamical mass generation through dynamical chiral symmetry breaking coincident with the removal of centre vortices, in contrast to the ASQTAD results \cite{Bowman:2010zr}. Notably, in the ultraviolet the vortex removed and untouched configurations reach similar values, suggesting we have kept short-range interactions intact through vortex removal. We attribute the small remaining dynamical mass generation to the ambiguity unavoidably present in maximal centre gauge fixing; since we have we selected a Gribov copy some centre vortex structure on the lattice may remain present. The renormalisation function on the vortex removed configurations show similar results; more tree-like behaviour in the infrared, evinced by a lessening of the suppression as compared to the untouched, and agreement in the ultraviolet. Combined, the mass and renormalisation functions show a vortex removed structure where dynamical chiral symmetry breaking is almost completely removed, while short-range behaviour remains.\par

\begin{figure}
\includegraphics[height=\PlotHeight]{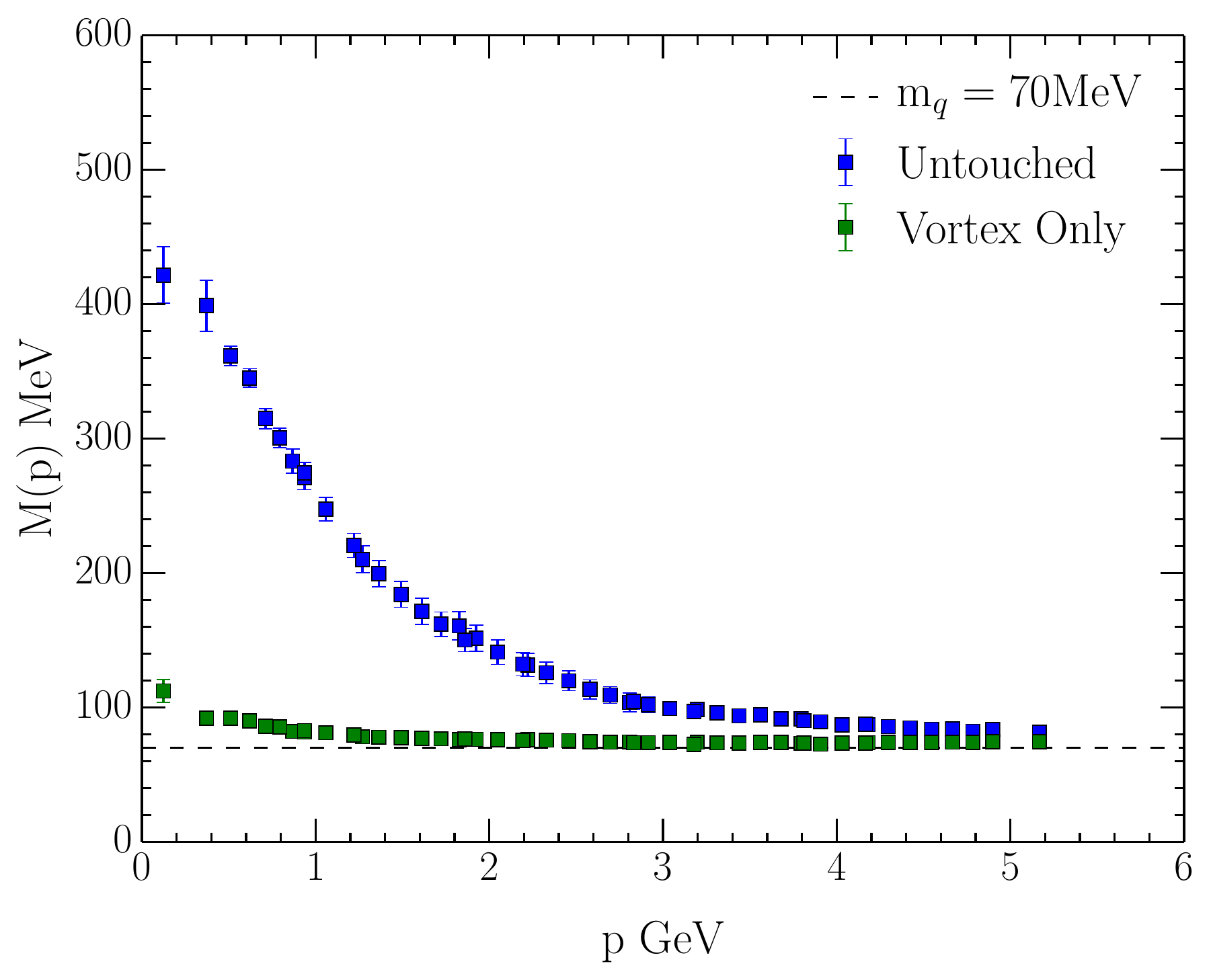}
\includegraphics[height=\PlotHeight]{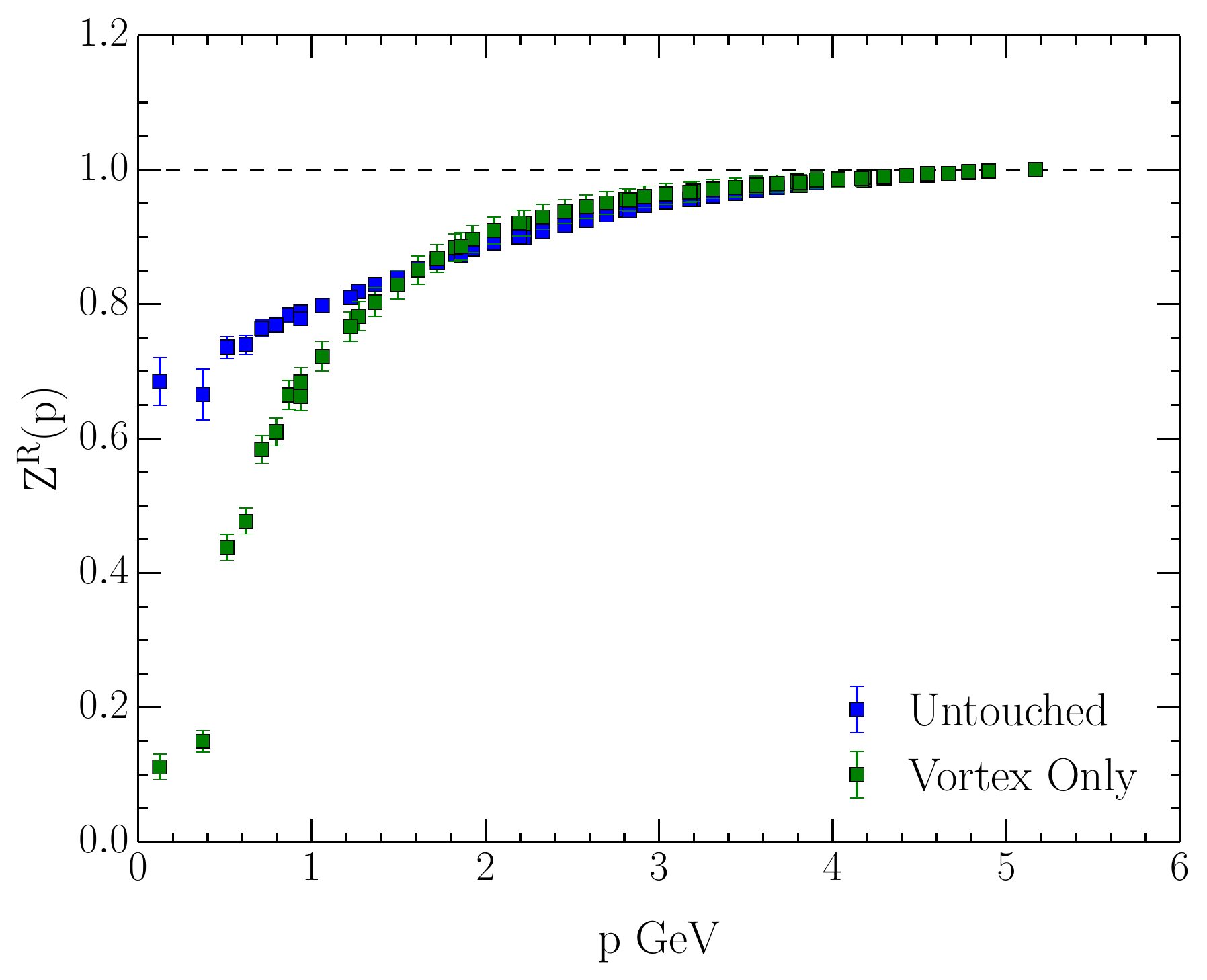}
\caption{The mass function and renormalisation functions on original (untouched) and vortex only configurations, at an input bare quark mass of $70$ MeV}
\label{fig:M02200UTVO}
\end{figure}
Vortex only results are plotted alongside untouched results in figure \ref{fig:M02200UTVO}. The mass function on vortex only configurations shows a complete absence of dynamical mass generation, with the mass function sitting on the input bare mass at all momenta. The vortex only background is not trivial, however, as can be seen from the renormalisation function, which dips sharply in the infrared, much more so than in the untouched case. This serves to suppress the quark propagator at long ranges, an indication that the vortex only background remains confining. \par
\section{The Overlap Quark Propagator on Cooled Backgrounds}
\begin{figure}
\includegraphics[height=\PlotHeight]{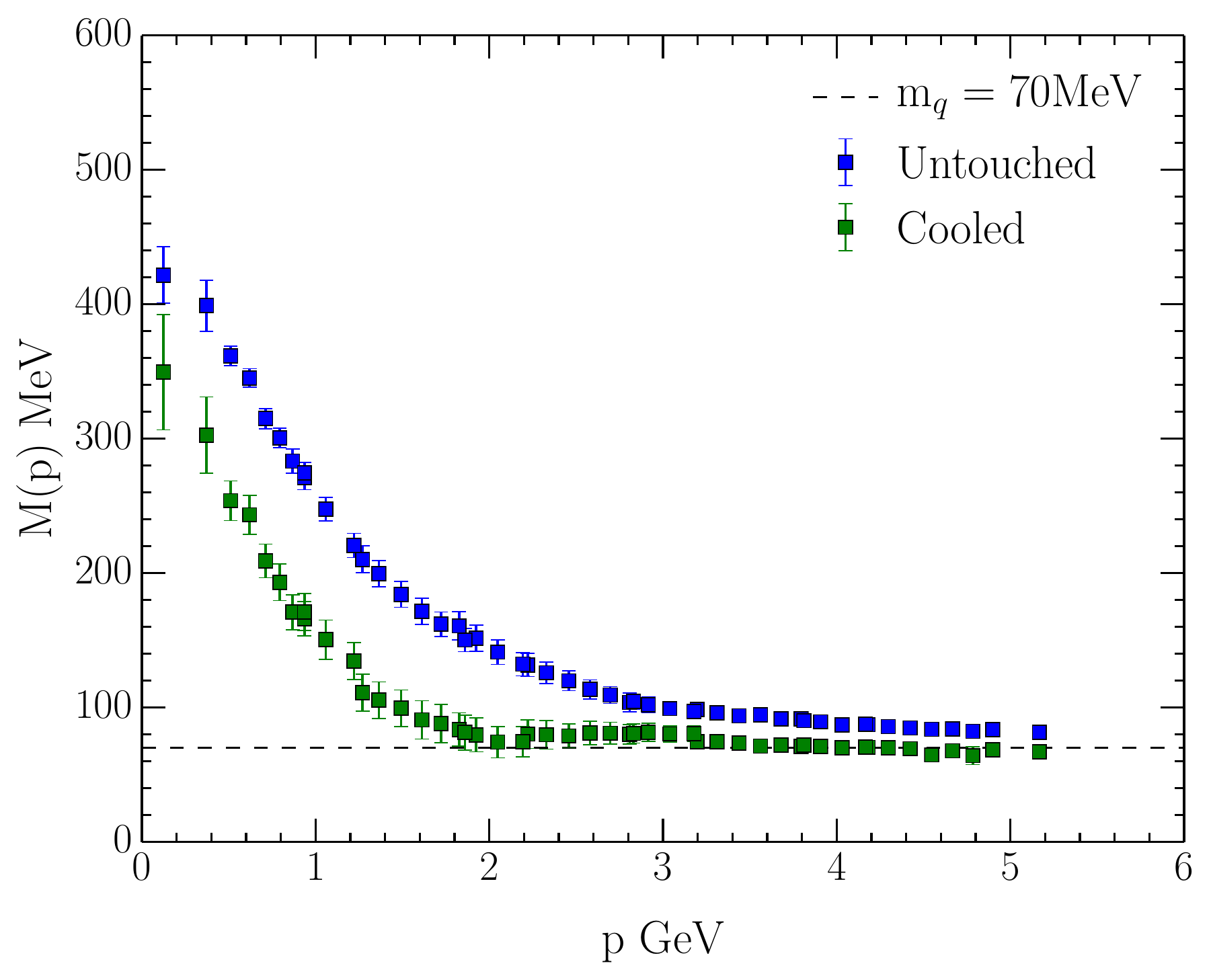}
\includegraphics[height=\PlotHeight]{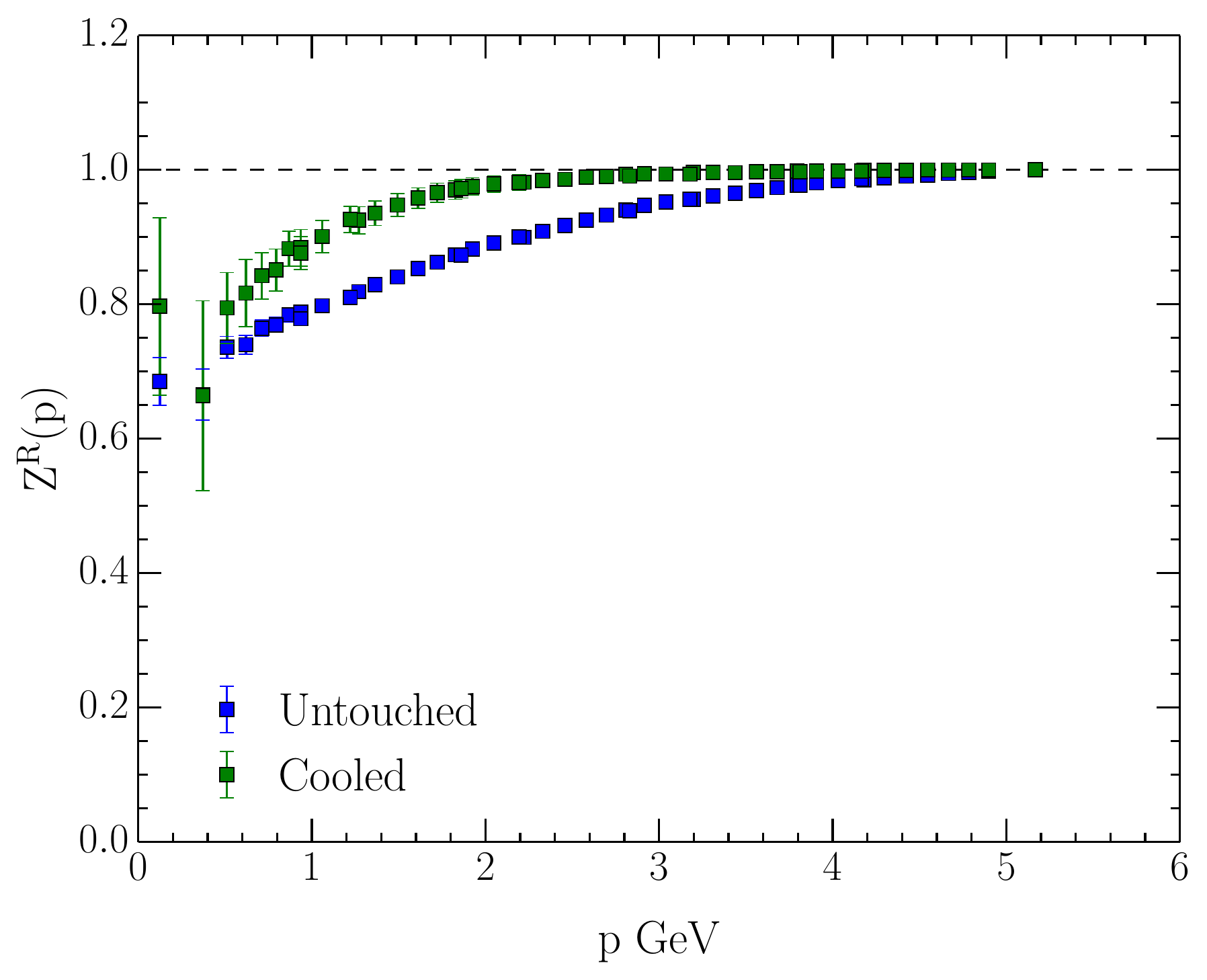}
\caption{The mass and renormalisation functions on original (untouched), at an input bare quark mass of $70$ MeV, with 40 sweeps of cooling}
\label{fig:M02200UTc}
\end{figure}

Due to the smoothness requirement of the overlap operator \cite{Narayanan:1994gw}, it is unclear to what extent it gives a true representation of the gauge field behaviour of very rough vortex only configurations. Thus, we perform cooling on vortex only configurations. 40 sweeps of cooling are performed using an $\mathcal{O}(a^4)$-three-loop improved action \cite{BilsonThompson:2002jk}. Under smearing, the overlap quark propagator retains its qualitative features at low momentum, with some loss of dynamical mass generation \cite{Trewartha:2013qga}, as can be seen in figure \ref{fig:M02200UTc} for the mass and renormalisation functions respectively. \par
\begin{figure}
\includegraphics[height=\PlotHeight]{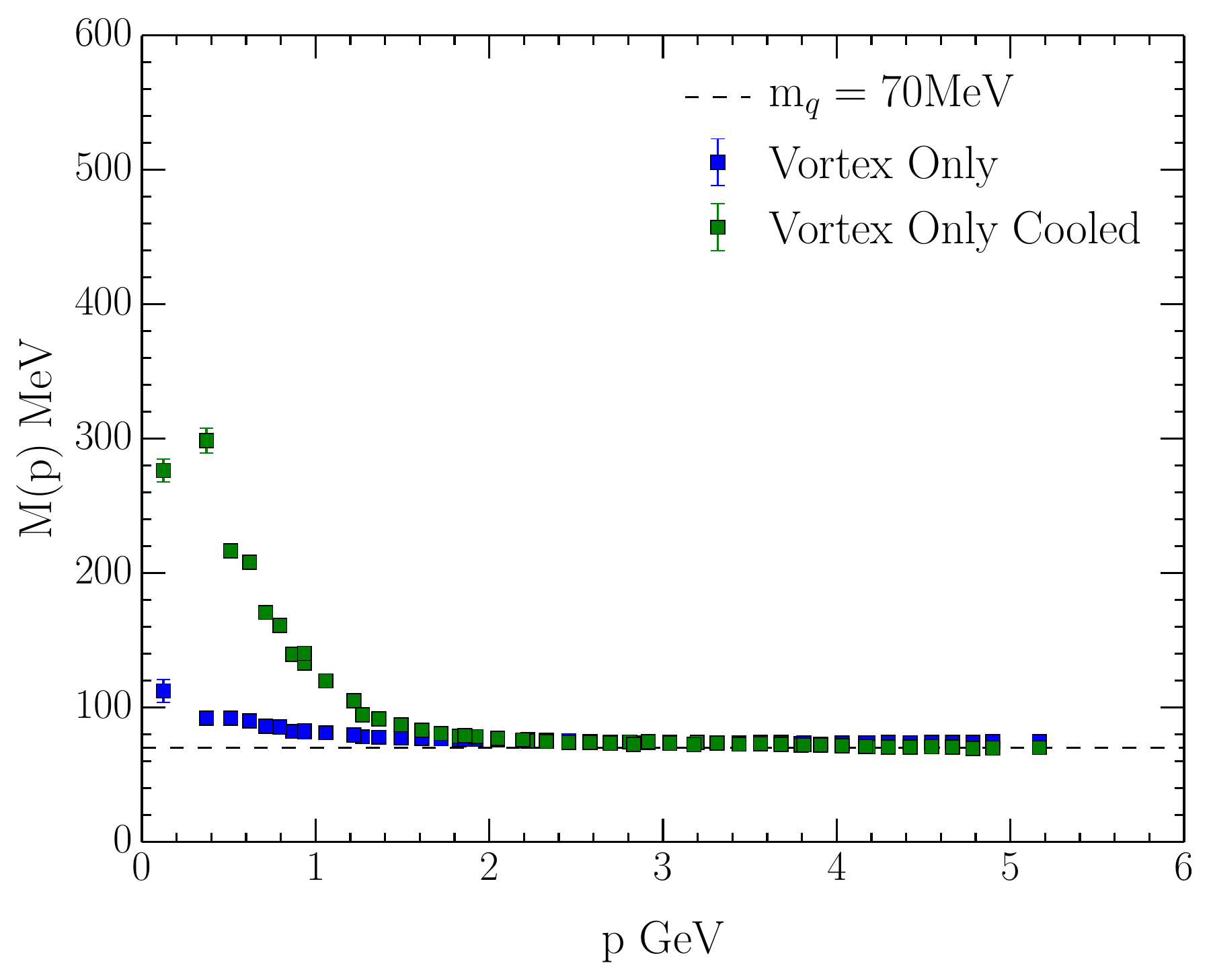}
\includegraphics[height=\PlotHeight]{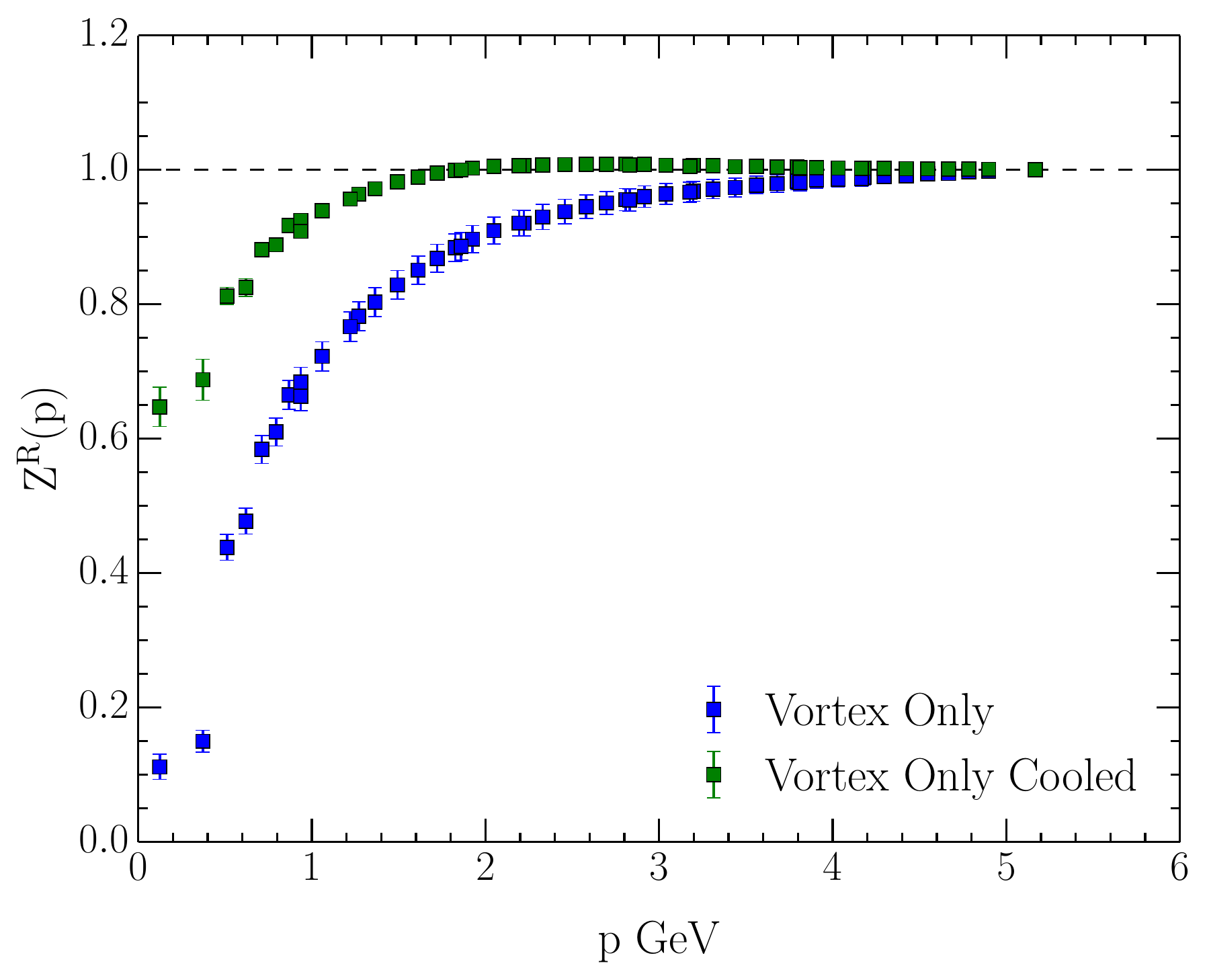}
\caption{The mass function and renormalisation functions on vortex only configurations, at an input bare quark mass of $70$ MeV, with 40 sweeps of cooling}
\label{fig:M02200VOc}
\end{figure}

\begin{figure}
\includegraphics[height=\PlotHeight]{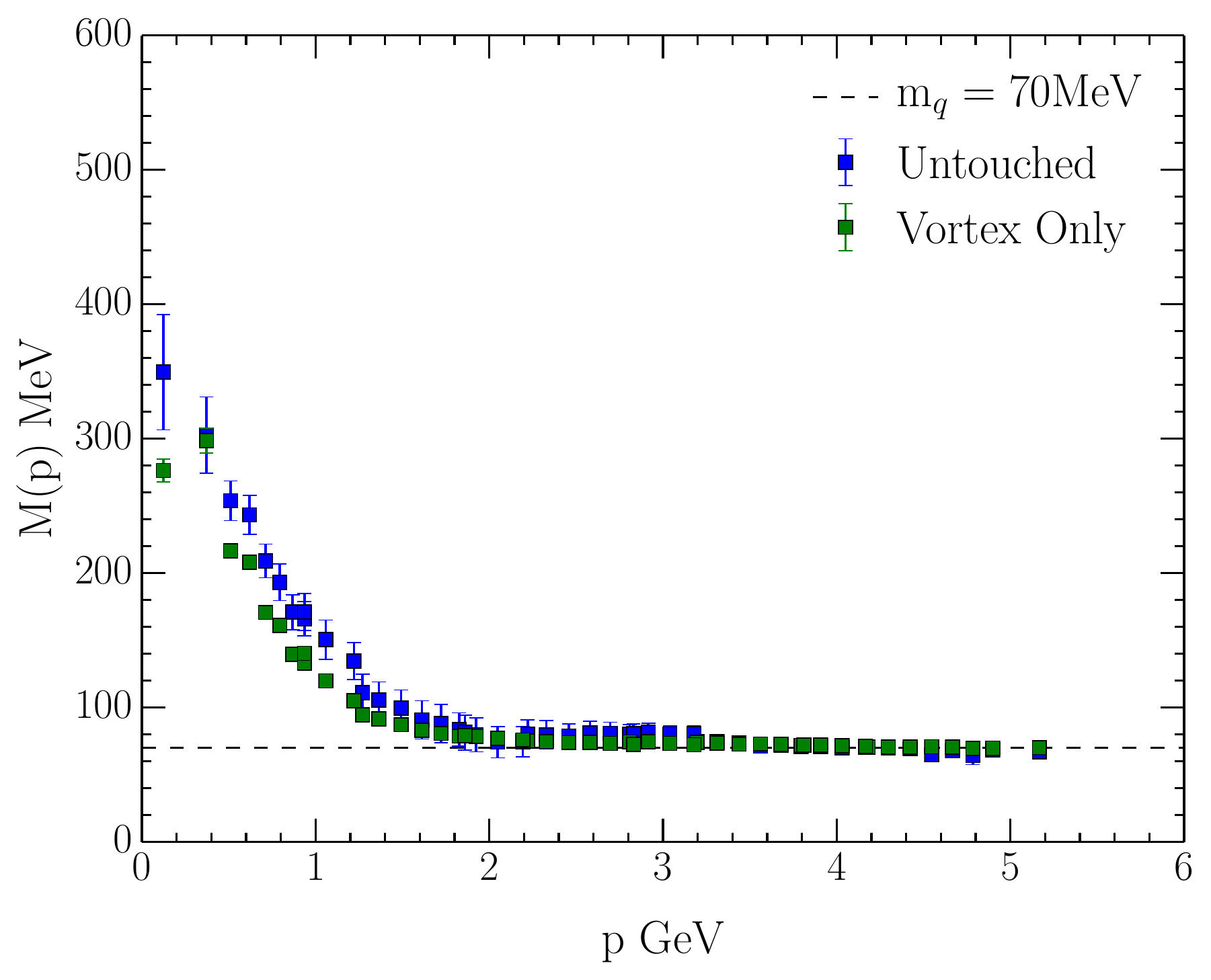}
\includegraphics[height=\PlotHeight]{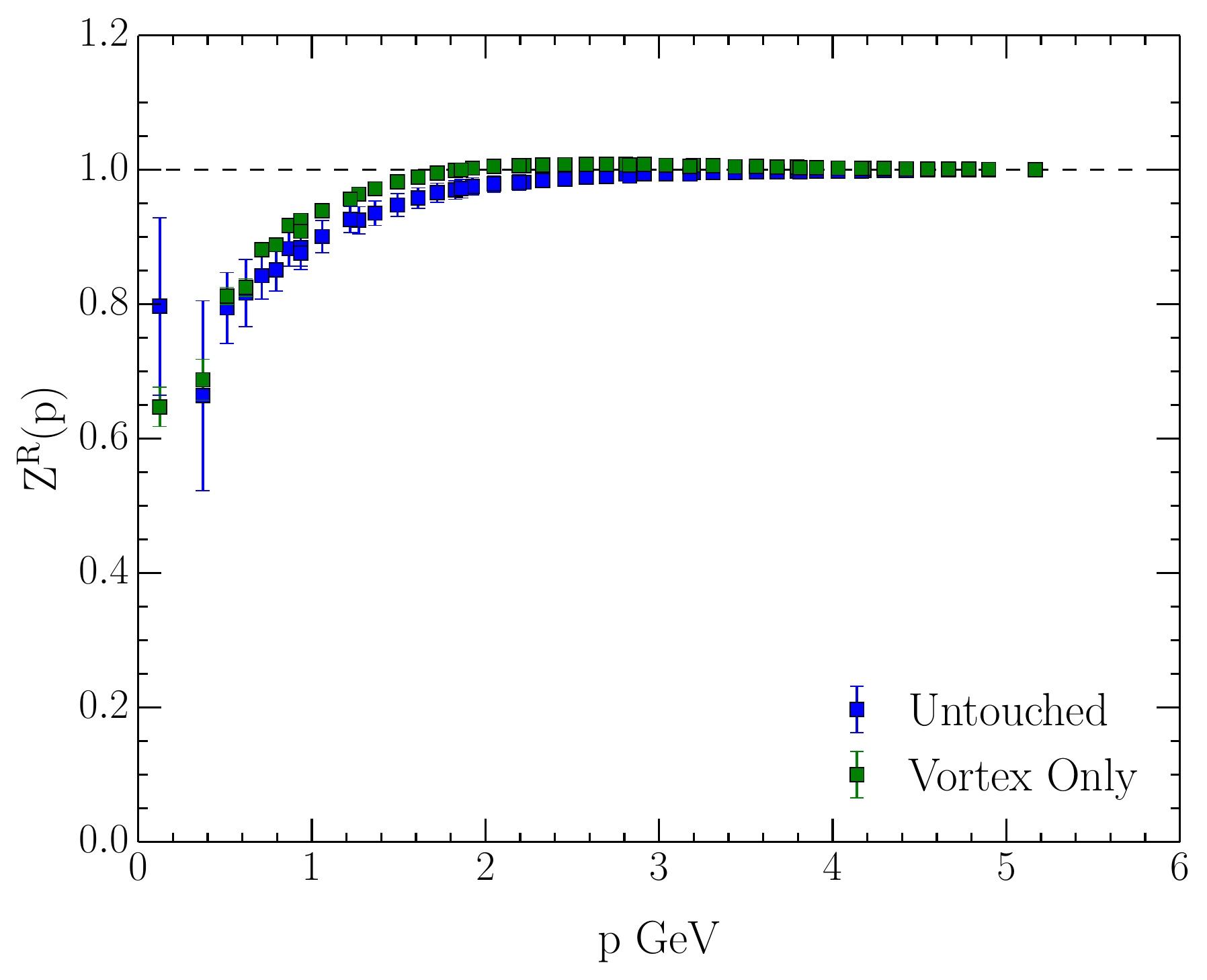}
\caption{The mass and renormalisation functions on vortex only and untouched configurations, at an input bare quark mass of $70$ MeV, with 40 sweeps of cooling}
\label{fig:M02200UTVOc}
\end{figure}

In the vortex only case, figure \ref{fig:M02200VOc}, there is a dramatic difference after cooling. On cooled configurations, there is a significant amount of dynamical mass generation present. The renormalisation function also dips far less dramatically in the ultraviolet. Combined, these two show a cooled vortex-only quark propagator with very similar behaviour to the untouched propagator, and hence a similar long range vacuum structure. This is particularly apparent when plotted side-by-side with the untouched results on cooled configurations, shown in figure \ref{fig:M02200UTVOc}. After cooling, vortex only results are almost identical to untouched results with a similar amount of cooling. Beginning from a background consisting solely of centre-vortices, we have been able to reproduce the long range features of the quark propagator.
\section{Conclusions}
We have created configurations with centre vortices removed, and configurations consisting solely of centre vortices, and investigated their chiral properties using the overlap quark propagator. We have been able to show for the first time loss of dynamical mass generation concurrent with vortex removal, suggesting a key role for centre vortices in dynamical chiral symmetry breaking in $SU(3)$, analogous to their role in $SU(2)$. Using cooling, we have been able to reproduce the long-range structure of the quark propagator on vortex only configurations, suggesting centre vortices are the key underlying long-range feature of the vacuum.
\acknowledgments
This research was undertaken on the NCI National Facility in Canberra, Australia, which is supported by the Australian Commonwealth Government. This research is supported by the Australian Research Council.


\begin{thebibliography}{31}
\bibitem{Del Debbio:1996mh} 
  L.~Del Debbio, M.~Faber, J.~Greensite and S.~Olejnik,
  Phys.\ Rev.\ D {\bf 55}, 2298 (1997)
  [hep-lat/9610005].
  
\bibitem{Del Debbio:1998uu} 
  L.~Del Debbio, M.~Faber, J.~Giedt, J.~Greensite and S.~Olejnik,
  Phys.\ Rev.\ D {\bf 58}, 094501 (1998)
  [hep-lat/9801027].
  
\bibitem{Greensite:2003bk} 
  J.~Greensite,
  Prog.\ Part.\ Nucl.\ Phys.\  {\bf 51}, 1 (2003)
  [hep-lat/0301023].
  
\bibitem{de Forcrand:1999ms} 
  P.~de Forcrand and M.~D'Elia,
  Phys.\ Rev.\ Lett.\  {\bf 82}, 4582 (1999)
  [hep-lat/9901020].
  
\bibitem{Langfeld:2003ev} 
  K.~Langfeld,
  Phys.\ Rev.\ D {\bf 69}, 014503 (2004)
  [hep-lat/0307030].
  
\bibitem{Gattnar:2004gx} 
  J.~Gattnar, C.~Gattringer, K.~Langfeld, H.~Reinhardt, A.~Schafer, S.~Solbrig and T.~Tok,
  Nucl.\ Phys.\ B {\bf 716}, 105 (2005)
  [hep-lat/0412032].
    
\bibitem{Bowman:2008qd} 
  P.~O.~Bowman, K.~Langfeld, D.~B.~Leinweber, A.~O' Cais, A.~Sternbeck, L.~von Smekal and A.~G.~Williams,
  Phys.\ Rev.\ D {\bf 78}, 054509 (2008)
  [arXiv:0806.4219 [hep-lat]].
  
\bibitem{Orginos:1999cr} 
  K.~Orginos {\it et al.}  [MILC Collaboration],
  Phys.\ Rev.\ D {\bf 60}, 054503 (1999)
  [hep-lat/9903032].
  
\bibitem{Bowman:2010zr} 
  P.~O.~Bowman, K.~Langfeld, D.~B.~Leinweber, A.~Sternbeck, L.~von Smekal and A.~G.~Williams,
  Phys.\ Rev.\ D {\bf 84}, 034501 (2011)
  [arXiv:1010.4624 [hep-lat]].
 
  
\bibitem{Narayanan:1992wx} 
  R.~Narayanan and H.~Neuberger,
  Phys.\ Lett.\ B {\bf 302}, 62 (1993)
  [hep-lat/9212019].
  
    %
\bibitem{Zanotti:2001yb} 
  J.~M.~Zanotti {\it et al.}  [CSSM Lattice Collaboration],
  Phys.\ Rev.\ D {\bf 65}, 074507 (2002)
  [hep-lat/0110216].
  
  
\bibitem{Leinweber:1998im} 
  D.~B.~Leinweber {\it et al.}  [UKQCD Collaboration],
  Phys.\ Rev.\ D {\bf 58}, 031501 (1998)
  [hep-lat/9803015].
  
\bibitem{Luscher:1984xn} 
  M.~Luscher and P.~Weisz,
  Commun.\ Math.\ Phys.\  {\bf 97}, 59 (1985)
  [Erratum-ibid.\  {\bf 98}, 433 (1985)].
  
\bibitem{Davies:1987vs} 
  C.~T.~H.~Davies, G.~G.~Batrouni, G.~R.~Katz, A.~S.~Kronfeld, G.~P.~Lepage, K.~G.~Wilson, P.~Rossi and B.~Svetitsky,
  Phys.\ Rev.\ D {\bf 37}, 1581 (1988).
  
\bibitem{Bonnet:1999mj} 
  F.~D.~R.~Bonnet, P.~O.~Bowman, D.~B.~Leinweber, A.~G.~Williams and D.~G.~Richards,
  Austral.\ J.\ Phys.\  {\bf 52}, 939 (1999)
  [hep-lat/9905006].
   
\bibitem{Narayanan:1994gw} 
  R.~Narayanan and H.~Neuberger,
  Nucl.\ Phys.\ B {\bf 443}, 305 (1995)
  [hep-th/9411108].
  
\bibitem{BilsonThompson:2002jk} 
  S.~O.~Bilson-Thompson, D.~B.~Leinweber and A.~G.~Williams,
  Annals Phys.\  {\bf 304}, 1 (2003)
  [hep-lat/0203008].
  
\bibitem{Trewartha:2013qga} 
  D.~Trewartha, W.~Kamleh, D.~Leinweber and D.~S.~Roberts,
  Phys.\ Rev.\ D {\bf 88}, 034501 (2013)
  [arXiv:1306.3283 [hep-lat]].
\end{thebibliography}
\end{document}